\documentclass[conference, a4paper]{IEEEtran}

\renewcommand\IEEEkeywordsname{Index Terms}

\usepackage{graphicx}
\usepackage{caption}
\usepackage{cite}
\usepackage[hyphens]{url}
\usepackage[nolist]{acronym} 
\usepackage{pgfplots}
\usepackage{tablefootnote}

\usepackage{tikz}
\usetikzlibrary{arrows,calc,shapes,graphs,graphs.standard,quotes,arrows.meta,decorations.markings,positioning,fit,positioning,hobby,decorations.text,decorations.pathreplacing}

\usepackage{pgfplots}
\pgfplotsset{compat=1.18}
\pgfdeclarelayer{background}
\pgfdeclarelayer{foreground}
\pgfsetlayers{background,main,foreground}
\usepgfplotslibrary{groupplots}

\pgfplotsset{compat=newest}
\usepackage[bookmarks=false]{hyperref}
\usepackage{amsmath, amsbsy, amssymb}
\usepackage{amsthm}
\usepackage{multirow}
\usepackage{nicematrix}
\usepackage{booktabs}
\usepackage{nicefrac}

\usepackage{etoolbox} %

\usepackage[linesnumbered,vlined, ruled]{algorithm2e}

\usepackage{etoolbox}
\makeatletter
\patchcmd\algocf@Vline{\vrule}{\vrule \kern-0.4pt}{}{}
\patchcmd\algocf@Vsline{\vrule}{\vrule \kern-0.4pt}{}{}
\makeatother

\usepackage{marginnote}
\tikzset{>=latex}
\SetAlCapNameFnt{\footnotesize}
\SetAlCapFnt{\footnotesize}

\definecolor{mittelblau}{RGB}{0, 126, 198}
\definecolor{violettblau}{cmyk}{0.9, 0.6, 0, 0}
\definecolor{rot}{RGB}{238, 28 35}
\definecolor{apfelgruen}{RGB}{140, 198, 62}
\definecolor{gelb}{RGB}{255, 229, 0}
\definecolor{orange}{RGB}{244, 111, 33}
\definecolor{pink}{RGB}{237, 0, 140}
\definecolor{lila}{RGB}{128, 10, 145}
\definecolor{hellgrau}{RGB}{224, 224, 224}
\definecolor{mittelgrau}{RGB}{128, 128, 128}
\definecolor{dunkelgrau}{RGB}{80,80,80}
\definecolor{anthrazit}{RGB}{19, 31, 31}
\definecolor{darkgreen}{RGB}{34,139,34}
\definecolor{aqua}{RGB}{0, 255, 255}

\definecolor{lightgray}{RGB}{211,211,211}

\definecolor{neuesgruen}{RGB}{61, 173, 65}
\definecolor{dunklereshellgrau}{RGB}{176, 176, 176}
\definecolor{neuesgelb}{RGB}{255,160,0}
\definecolor{neuescyan}{RGB}{69,185,224}

\definecolor{tollesgruen}{RGB}{0,217,171}
\definecolor{tollesmagenta}{RGB}{197,67,143}

\definecolor{tollesgelb}{RGB}{255,199,95}
\definecolor{tollesrot}{RGB}{255,111,145}

\colorlet{R12}{apfelgruen}
\colorlet{R23}{mittelblau}
\colorlet{R45}{pink}

\tikzset{
       vnd/.style={
        shape=circle,
        fill=black,
        draw,
        inner sep=0pt,
        minimum size=0.2cm},
        cnd/.style={
        shape=rectangle,
        fill=white,
        draw,
        minimum width=0.05mm,
        minimum height = 0.05mm}, 
         vndR/.style={
        shape=circle,
        fill=red,
        draw,
        inner sep=0pt,
        minimum size=0.2cm},
        cndR/.style={
        shape=rectangle,
        fill=white,
        draw=red,
        minimum width=0.05mm,
        minimum height = 0.05mm}
}

\IEEEoverridecommandlockouts

\renewcommand{\vec}[1]{\boldsymbol{#1}}

\newcommand{\Hm}{\vec{H}}

\newcommand{\LB}{\left(}
\newcommand{\RB}{\right)}

\renewcommand{\ln}[1]{\mathop{\mathrm{ln}}\LB #1\RB}

\begin{document}

\begin{NoHyper}
\title{Long Polar vs. LDPC Codes under Complexity-Constrained Decoding}

\author{\IEEEauthorblockN{Felix Krieg, Marvin Rübenacke, Andreas Zunker, and Stephan ten Brink\\}
	\IEEEauthorblockA{
		Institute of Telecommunications, Pfaffenwaldring 47, University of  Stuttgart, 70569 Stuttgart, Germany 
		\\\{krieg, ruebenacke, zunker, tenbrink\}@inue.uni-stuttgart.de\\
	}
		\thanks{This work is supported by the German Federal Ministry of Research, Technology and Space (BMFTR) within the project Open6GHub (grant no. 16KISK019).} 
  }

\makeatletter
\patchcmd{\@maketitle}  %
{\addvspace{0.5\baselineskip}\egroup}
{\addvspace{-1.8\baselineskip}\egroup}
{}
{}
\makeatother

\maketitle

\begin{acronym}
\acro{ML}{maximum likelihood}
\acro{BP}{belief propagation}
\acro{BPL}{belief propagation list}
\acro{LDPC}{low-density parity-check}
\acro{BER}{bit error rate}
\acro{SNR}{signal-to-noise-ratio}
\acro{BPSK}{binary phase shift keying}
\acro{AWGN}{additive white Gaussian noise}
\acro{LLR}{log-likelihood ratio}
\acro{MAP}{maximum a posteriori}
\acro{FER}{frame error rate}
\acro{BLER}{block error rate}
\acro{SCL}{successive cancellation list}
\acro{SCAL}{successive cancellation automorphism list}
\acro{SC}{successive cancellation}
\acro{BI-DMC}{Binary Input Discrete Memoryless Channel}
\acro{CRC}{cyclic redundancy check}
\acro{CA-SCL}{CRC-aided successive cancellation list}
\acro{PAC}{polarization-adjusted convolutional}
\acro{BEC}{Binary Erasure Channel}
\acro{BSC}{Binary Symmetric Channel}
\acro{BCH}{Bose--Ray-Chaudhuri--Hocquenghem}
\acro{RM}{Reed--Muller}
\acro{RS}{Reed-Solomon}
\acro{SISO}{soft-in/soft-out}
\acro{3GPP}{3rd Generation Partnership Project }
\acro{eMBB}{enhanced Mobile Broadband}
\acro{CN}{check node}
\acro{VN}{variable node}
\acro{GenAlg}{Genetic Algorithm}
\acro{CSI}{Channel State Information}
\acro{OSD}{ordered statistic decoding}
\acro{MWPC-BP}{minimum-weight parity-check BP}
\acro{FFG}{Forney-style factor graph}
\acro{MBBP}{multiple-bases belief propagation}
\acro{URLLC}{ultra-reliable low-latency communications}
\acro{mMTC}{massive machine-type communications}
\acro{DMC}{discrete memoryless channel}
\acro{SGD}{stochastic gradient descent}
\acro{QC}{quasi-cyclic}
\acro{5G}{fifth generation mobile telecommunication}
\acro{SCAN}{soft cancellation}
\acro{LSB}{least significant bit}
\acro{MSB}{most significant bit}
\acro{AED}{automorphism ensemble decoding}
\acro{AE-SC}{automorphism ensemble successive cancellation}
\acro{PPV}{Polyanskyi--Poor--Verd\'{u}}
\acro{RREF}{reduced row echelon form}
\acro{PL}{permutation linear}
\acro{LTA}{lower triangular affine}
\acro{BLTA}{block lower triangular affine}
\acro{PTPC}{pre-transformed polar codes}
\acro{UPO}{universal partial order}
\acro{SSC}{simplified successive cancellation}
\acro{SSCL}{simplified successive cancellation list}
\acro{ASIC}{application-specific integrated circuit}
\acro{LMS}{layered min-sum}
\acro{HARQ}{hybrid automatic repeat request}
\acro{DE}{density evolution}
\end{acronym}

\begin{abstract}
The prevailing opinion in industry and academia is that polar codes are competitive for short code lengths, but can no longer keep up with \ac{LDPC} codes as block length increases.
This view is typically based on the assumption that \ac{LDPC} codes can be decoded with a large number of  \ac{BP} iterations. However, in practice, the number of iterations may be rather limited due to latency and complexity constraints.
In this paper, we show that for a similar number of fixed-point \ac{LLR} operations, long polar codes under \ac{SC} decoding outperform their \ac{LDPC} counterparts. In particular, \ac{SSC} decoding of polar codes exhibits a better complexity scaling than \textit{N}\,log\,\textit{N} and requires fewer operations than a single \ac{BP} iteration of an \ac{LDPC} code with the same parameters.
\end{abstract}
\acresetall


\section{Introduction}
\Ac{LDPC} codes are today's workhorse of error correction \cite{nozaki2022ldpc}.
Their well-understood, iterative \ac{BP} decoder and near-Shannon-limit performance  \cite{chung2001ldpcshannon} made \ac{LDPC} codes the preferred choice for numerous communication standards, including IEEE 802.3 10G Ethernet, IEEE 802.11 WiFi, IEEE 802.16 WiMAX, DVB-S/T/C2, CCSDS, and the 5G NR data channel.
While \ac{LDPC} codes closely approach the Shannon limit in practice, polar codes were the first to be provably capacity-achieving as the block length tends to infinity \cite{ArikanMain}.
Nonetheless, it was the development of short-length polar coding techniques---such as the incorporation of an outer \ac{CRC} code and the introduction of the \ac{SCL} decoding algorithm \cite{talvardyList}---that established polar codes as a leading solution in the short block length regime and finally led to their inclusion in the 5G NR control channel \cite{polar5G2018}.
Since their introduction in the standard, further progress has been made in academia. Different pre-transformations \cite{arikan2019pac, zunker2025rowmerged} enable further performance gains, while \ac{SSC} \cite{simplifiedSC} and \ac{SSCL} \cite{hashemi2017flexfastsscl} allow for even more efficient decoder implementations.
In consideration of the upcoming 6G standard, the already considerable performance in \ac{5G}, and the potential incorporation of those recent advances, polar codes emerge as a promising candidate for a second standardization.

In addition to ``raw'' error-rate performance, however, another design objective of a new mobile communications standard could be the unification of decoding schemes \cite{bits2023unified}. Rather than employing a variety of codes for each use case, it is beneficial to implement a single code with a standardized, simplified description. This approach would conserve resources, both during the implementation phase and subsequently, in the deployed hardware. However, for such a goal to be realized, the code employed must exhibit sufficient performance and energy efficiency over a broad spectrum of lengths and rates. In the context of 6G as the successor to 5G, the question therefore arises: is it possible to design \ac{LDPC} codes that achieve the desired performance even at short lengths, or can satisfactory results also be achieved with polar codes at long lengths?

In recent years, numerous efforts have focused on improving the performance of short-length \ac{LDPC} codes (see, e.g., \cite{krieg2025ensemble} and references therein; also \cite{mandelbaum2025sced}, \cite{rosseel2022shortldpc}), whereas the alternative approach of employing long polar codes has received relatively little attention, with only a few exceptions such as \cite{liu2018516gbpsdecoderasicpolar}, \cite{yang2025perturbation}.

In our opinion, there are three prevalent misconceptions about polar codes which have led to the perception that \ac{LDPC} codes are superior in the long block length regime:
\begin{enumerate}
    \item The larger scaling exponent of polar codes \cite{korada2009scaling, mondelli2015errorexponent} compared to \ac{LDPC} codes make them operate far from the channel capacity for finite lengths;
    \item For polar codes to be competitive at all, highly complex decoding algorithms such as list or ensemble decoding are required;
    \item The $\mathcal{O}(N \log N)$ complexity scaling of polar encoding and decoding is prohibitively large compared to the linear scaling of \ac{LDPC} codes.
\end{enumerate}
However, the evolution of mobile communication standards has given rise to challenges in energy efficiency, and as a result, \ac{BP} decoders with large number of iterations---despite their superior error correction capabilities---may no longer be considered optimal \cite{kestel2018implementationwall}.
To the best of our knowledge, the literature lacks a comprehensive comparison between long polar codes and \ac{LDPC} codes.
In this paper, we want to show that the choice between polar and \ac{LDPC} codes is less clear-cut than commonly assumed.
To this end, we show using both theoretic analysis and Monte Carlo simulation that polar codes are competitive against \ac{LDPC} codes in the long block length regime, especially under decoding complexity constraints.
Our key findings are summarized as follows and directly address the misconceptions above:
\begin{enumerate}
    \item While asymptotically the error rate of polar codes does not decrease as fast as that of \ac{LDPC} codes, their performance under finite decoding complexity is similar;
    \item This competitive performance of polar codes can be achieved using low-complexity \ac{SSC} decoding;
    \item \Ac{SSC} decoding of polar codes scales better than $\mathcal{O}(N \log N)$. For any practical codeword length, the additional factor of $\log N$ is smaller than the multiplicative constants of \ac{BP}-based \ac{LDPC} decoding. Most striking, a full polar decoding is significantly less complex than a single iteration of \ac{BP}.
\end{enumerate}

\section{Preliminaries}

\subsection{Polar Codes \& Successive Cancellation Decoding}
Polar codes are constructed using the polar transform, defined by the $N \times N$ Hadamard matrix $\boldsymbol{G}_N = \left[\begin{smallmatrix} 1 & 0 \\ 1 & 1 \end{smallmatrix}\right]^{\otimes n}$, where $N = 2^n$ and $(\cdot)^{\otimes n}$ denotes the $n$-th Kronecker power.
This transform polarizes a set of identical channels into synthetic channels that are either highly reliable or highly unreliable.
Based on the code design, $K$ of the most reliable channels are selected to carry information; their indices form the \emph{information set} $\mathcal{I}$.
The remaining, unreliable channels constitute the \emph{frozen set} $\mathcal{F}$.
The generator matrix $\boldsymbol{G}$ of the polar code is formed by selecting the rows of $\boldsymbol{G}_N$ corresponding to the indices in $\mathcal{I}$.
As the block length $N$ of polar codes are inherently limited to powers of 2, length-matching techniques such as puncturing and shortening must be applied to achieve flexibility in $N$ \cite{nui2013lengthmatching}.

Polar codes were first introduced with the \ac{SC} decoding algorithm \cite{ArikanMain}.
This algorithm determines the most likely value for each information bit $u_i$ in a successive order, i.e.,
\begin{equation*}
    \hat{u}_i = \arg \max_{u_i} \operatorname{Pr}\{u_i | \hat{u}_0 \dots \hat{u}_{i-1}, \boldsymbol{y}\},
\end{equation*}
where $\boldsymbol{y}$ denotes the received vector. For frozen bits $i\in \mathcal{F}$, $\hat{u}_i = 0$.
\Ac{SC} decoding can be formulated recursively by splitting the channel \ac{LLR} vector in two halves $\boldsymbol{L} = (\boldsymbol{L}_1 \mid \boldsymbol{L}_2) $. First, the \ac{LLR} for the first half is computed as
\begin{equation*}\label{eq:scg}
    L_{1,i}' = L_{1,i} \boxplus L_{2,i} \approx \operatorname{sgn}(L_{1,i})\operatorname{sgn}(L_{2,i}) \min \{ |L_{1,i}|, |L_{2,i}| \},
\end{equation*}
where 
$a \boxplus b = \ln {\frac{1+\operatorname{e}^{a+b}}{\operatorname{e}^{a}+\operatorname{e}^{b}}}$.
Then, $\boldsymbol{L}_1'$ is decoded further recursively, resulting in the estimate $\hat{\boldsymbol{x}}_1'$, which is used to compute the \ac{LLR} for the second half as
\begin{equation*}
    L_{2,i}' = (-1)^{\hat{x}_{1,i}'} L_{1,i} + L_{2,i},
\end{equation*}
and the decoding of $\boldsymbol{L}_2'$ is performed recursively, resulting in $\hat{\boldsymbol{x}}_2'$. The final codeword estimate is then given as 
\begin{equation*}
    \hat{\boldsymbol{x}}=(\hat{\boldsymbol{x}}_1'\oplus\hat{\boldsymbol{x}}_2' \mid \hat{\boldsymbol{x}}_2').
\end{equation*}
While the recursion can be continued all the way to the information bits $u_i$, practical polar code designs (i.e., information sets $\mathcal{I}$) result in simple to decode leaf nodes of larger size $s>1$ \cite{simplifiedSC}:
\begin{enumerate}
    \item Rate-0: $\hat{\boldsymbol{x}}=\boldsymbol{0}$ independently of $ \boldsymbol{L}$;
    \item Repetition: $\hat{x}_i = h(\sum_{j=0}^{s-1}L_j)$;
    \item Single parity-check: $\hat{x}_i = h(L_i)$ with the bit in the position of the smallest $|L_i|$ flipped if $\sum_{i=0}^{s-1} h(L_i) \not\equiv 0 \mod 2$;
    \item Rate-1: $\hat{x}_i = h(L_i)$;
\end{enumerate}
where $h(\cdot)$ denotes the hard decision function on an \ac{LLR}.

\subsection{LDPC Codes \& Belief Propagation Decoding}
\Ac{LDPC} codes are linear codes characterized by a sparse parity-check matrix, $\Hm$ \cite{Gallager, mackay99bp}.
An alternative equivalent representation of the code is its Tanner graph \cite{tanner}.
The Tanner graph is a bipartite graph, consisting of \acp{CN} (corresponding to the parity check equations) and \acp{VN} (corresponding to the codeword bits), where there exists an edge between \ac{CN} $i$ and \ac{VN} $j$ if and only if $h_{i,j} = 1$. The number of adjacent \acp{VN} to \ac{CN} $i$ defines the \ac{CN} degree $d_{\mathrm{CN},i}$. 

Efficient decoding of \ac{LDPC} codes is achieved by the \ac{BP} algorithm, which leverages the sparse connectivity of the Tanner graph.
In \ac{BP}, messages between \acp{CN} and \acp{VN} and vice versa are passed iteratively along the edges of the graph.
The update equation for the \ac{VN} $j$ is 
\begin{equation*}
    q_{j\to i} = L_{\mathrm{ch},j} + \sum_{i'\neq i}r_{i'\to j},
\end{equation*}
where $L_{ch,j}$ is the channel \ac{LLR} and $r_{i\to j}$ is the incoming message from \ac{CN} $i$.
Similarly, the update for the \ac{CN} $i$ is 
\begin{equation}\label{eq:spa_cn}
    r_{i \to j} = 2\cdot \tanh^{-1}\Bigg( \prod_{j'\neq j} \tanh \left(\frac{ q_{j'\to i}}{2} \right) \Bigg).
\end{equation}
In the first iteration, the messages are initialized with the channel \acp{LLR} as no other (extrinsic) information is available. Thus, for the initialization $q_{j\to i}=L_{\mathrm{ch}, j}$.
After $N_\mathrm{it}$ iterations the final output is calculated by 
\begin{equation*}
    Q_{j} = L_{\mathrm{ch},j} + \sum_{i} r_{i \to j}.
\end{equation*}

In practice, two modifications are done to allow for efficient implementations of the \ac{BP} algorithm.
First, the \ac{CN} function is replaced by a hardware-friendly approximation computing the minimum of the absolute values of the incoming messages, which dominates the value in \eqref{eq:spa_cn}. 
However, this overestimates the \ac{CN} output, and therefore, an attenuation factor of $\alpha \approx 0.75$ is introduced \cite{chen2002minsum}.
Second, instead of updating all \acp{CN} and \acp{VN} at once, a layered schedule is performed.
Let $\mathcal{R}$ denote the set of all \acp{CN}. In a layered schedule, only a subset of \acp{CN} $\mathcal{R}_i \subseteq \mathcal{R}$ of size $N_\mathrm{L}$ is updated at a time, followed by the update of their connected \acp{VN}. An iteration is complete when all \acp{CN} have been updated exactly once.
Typically, layered decoding reduces the number of required decoding iterations by the factor two, however, at an increased latency due to the sequential processing schedule \cite{hocevarlayered}\cite{Litsyn_Schedules_LDPC}.
The resulting \ac{LMS} decoder stores the \ac{CN} output messages $r_{j\to i}$ and the total \ac{LLR} $Q_j$ of each \ac{VN} in memory, which are initialized to ${r_{j\to i}^{(0)}=0}$ and ${Q_j^{(0)}= L_{\mathrm{ch},j}}$. 

The \ac{VN}-to-\ac{CN} messages are computed on the fly as
\begin{equation}\label{eq:msq}
    q_{i\to j} = Q_j^{(t)} - r_{j\to i}^{(t)}.
\end{equation}
These intermediate values are then used to compute $r_{j\to i}$ in the \ac{CN} update as
\begin{equation*}
    r_{j\to i}^{(t+1)} = \alpha \cdot \Bigg( \prod_{j' \ne j} \operatorname{sgn}(q_{i\to j'}) \Bigg) \cdot \min_{j'\ne j} {|q_{i\to j'}|}
\end{equation*}
and the total \ac{LLR} of the \ac{VN} as
\begin{equation}\label{eq:msvn}
    Q_j^{(t+1)} = q_{i\to j} + r_{j\to i}^{(t+1)}.
\end{equation}

\section{Complexity of LDPC and Polar Decoding}

\subsection{General Considerations}
In order to estimate the decoding complexity, a conservative estimate is presented based on the required elementary \ac{LLR} operations.
We assume that these operations dominate the computational complexity compared to bit operations, as at least 4 to 5 bits are required to accurately represent the \acp{LLR} for both \ac{SC} and \ac{LMS} decoding \cite{shi2014quantizedsc, zhao2005quantizedminsum}.
While the number of \ac{LLR} operations clearly cannot solely reflect the efficiency of \ac{ASIC} implementation which also has to consider parallelism, locality, interconnects and memory \cite{kestel2018implementationwall}, it can serve as a preliminary indicator of the energy consumption of the respective decoders.

For both polar \ac{SSC} and \ac{LDPC} \ac{LMS} decoders, the main \ac{LLR} computations are additions/subtractions (ADD) and minimum (MIN) operations.
Moreover, a minimum operation compares two values and, thus, is in its essence a subtraction.
Therefore, it is reasonable to assume that both ADD and MIN operations constitute similar computational complexity.

\subsection{Min-Sum Decoding of LDPC Codes}
We can compute the required number of equivalent operations for the \ac{LMS} decoding of an \ac{LDPC} code as
\begin{equation*}
    N_{\mathrm{Ops,{LMS}}} = N_\mathrm{iter}\cdot \bigg( 5 \cdot \sum_{i,j} h_{i,j} - 3 M \bigg),
\end{equation*}
where $N_\mathrm{iter}$ denotes the number of \ac{BP} iterations.
The constant factor of $5$ for each edge in the Tanner graph originates from
\begin{itemize}
    \item two binary additions ((\ref{eq:msq}) and (\ref{eq:msvn}))
    \item one addition to account for the multiplication with $\alpha$ in the attenuated Min-Sum update in \eqref{eq:msq}, which is efficiently implemented by subtracting a bit-shifted version of itself
    \item two MIN operations in the \ac{CN} step, \eqref{eq:msq}.
\end{itemize}
Note that for each \ac{CN}, the smallest and the second smallest absolute value have to be found. Thus,  $d_\mathrm{CN}-1$ MIN operations have to be performed for the first minimum, and $d_\mathrm{CN}-2$ MIN operations to find the second minimum, which is 3 less than the number of adjacent edges. Thus, we subtract $3M$.

\subsection{Successive Cancellation Decoding of Polar Codes}

For the polar code of length $N=2^n$, we can calculate the number of equivalent operations under plain \ac{SC} decoding to be $N/2$ MIN operations and $N/2$ ADD operations per stage. Since there are $\log_2(N)=n$ stages, we can find the operations to be 
\begin{equation*}
    N_{\mathrm{Ops,SC}} = N\cdot\log_2(N).
\end{equation*}
Note that, in contrast to \ac{LDPC} codes, no correction factor for the \ac{CN} computation is required to closely match the performance of decoding with the ``$\boxplus$''-function. 
For \ac{SSC}, depending on the code design $\mathcal{I}$, the recursion terminates already at larger leaf nodes. While rate-0 and rate-1 nodes do not require any computation, repetition nodes involve $s-1$ ADD operations, while single parity-check nodes require $s-1$ MIN operations to determine the position of the bit to be flipped.
Unfortunately there is no closed-form expression for $N_{\mathrm{Ops,SSC}}$, but it can be computed recursively given $\mathcal{I}$.
While clearly $N_{\mathrm{Ops,SSC}}\le N_{\mathrm{Ops,SC}}$, there does not exist a theoretic analysis of its scaling yet.
Note that the decoding latency (i.e., the number of successive steps required to decode) was proven to be being sub-linear for \ac{SSC} \cite{mondelli2021sublinearlatency} compared to the linear $\mathcal{O}(N)$ scaling of conventional \ac{SC} decoding.

Fig.~\ref{fig:complexityperinfobit} shows the computational complexity per information bit of \ac{SC} decoding and simplified \ac{SC} decoding. We observe that simplified \ac{SC} decoding does not only reduce the complexity by a multiplicative factor, but changes the scaling law to sub-$\mathcal{O}(N\log N)$, as the curves are no longer straight lines in the normalized, logarithmic plot.

\section{Numerical Results}
To enable a fair comparison between the two code families, we consider codes with varying lengths, rates, and designs.
The specific parameters of the codes examined in this study are summarized in Table~\ref{tab:params}, including the payload size $K$, codeword length $N$, and code rate $R=\nicefrac{K}{N}$.

\begin{table}[htb]
    \centering
    \caption{\footnotesize The considered code designs and parameters.}
    \label{tab:params}
    \begin{NiceTabular}{l|cccc}
    \toprule
    Polar design       & \multicolumn{4}{c}{Density evolution @ BLER $10^{-6}$} \\ 
    LDPC design        & \multicolumn{3}{c}{CCSDS} & DVB-S2\tablefootnote{{The DVB-S2 standard employs an outer \ac{BCH} code which is required for satisfactory performance. 
    For a direct comparison, we do not consider the decoding complexity of the \ac{BCH} code and solely focus on the inner \ac{LDPC} code. For the \ac{BLER} simulations, the \ac{BCH} code was applied.}} \\ 
    \cmidrule(r){1-1}\cmidrule(lr){2-4}\cmidrule(lr){5-5}
    Dimension $K$ & $1\,024$ & $4\,096$ & $16\,384 $ & $N\cdot R$        \\ 
    Length $N$    & \multicolumn{3}{c}{$\nicefrac{K}{R}$} & $64\,800$ \\
    \midrule
    Code Rate $R$ & \multicolumn{4}{c}{$\nicefrac{1}{2},\,\nicefrac{2}{3},\,\nicefrac{4}{5}$} \\
    \bottomrule
    \end{NiceTabular}
\end{table}

\begin{table*}[htb]
    \centering
    \vspace{1mm}
    \caption{\footnotesize Computational complexity in terms of \ac{LLR} operations per information bit. The \ac{LDPC} decoder uses early stopping and an iteration count to match the performance of the polar code at a \ac{BLER} of $10^{-3}$.}
    \label{tab:complexity}
    
    \setlength{\tabcolsep}{4pt}
    \begin{NiceTabular}{l|ccc|ccc|ccc|ccc}
    \toprule
     & \multicolumn{3}{c}{$K=1024$} & \multicolumn{3}{c}{$K=4096$} & \multicolumn{3}{c}{$K=16384$} & \multicolumn{3}{c}{$N=64800$} \\ 
    \midrule
    Code rate $R$ & $\nicefrac{1}{2}$ & $\nicefrac{2}{3}$ & $\nicefrac{4}{5}$
    & $\nicefrac{1}{2}$ & $\nicefrac{2}{3}$ & $\nicefrac{4}{5}$ & $\nicefrac{1}{2}$ & $\nicefrac{2}{3}$ & $\nicefrac{4}{5}$
    & $\nicefrac{1}{2}$ & $\nicefrac{2}{3}$ & $\nicefrac{4}{5}$ \\ 
    \midrule
    Polar \ac{SSC} $N_{\mathrm{Ops,SSC}} / K$ 
    & \phantom{0}13.85 & \phantom{0}9.76 & \phantom{0}7.34 & \phantom{0}15.87 & \phantom{0}11.66 & \phantom{00}8.71 & \phantom{0}17.68 & \phantom{0}13.01 & \phantom{00}9.93 & \phantom{0}18.26 & \phantom{0}13.88 & \phantom{0}10.99 \\
    LDPC \ac{LMS} $N_{\mathrm{Ops,LMS}} / K$ 
    & 191.40 & 94.34 & 75.35 & 236.15 & 132.50 & 102.07 & 334.36 & 178.43 & 135.85 & 320.00 & 234.26 & 108.75 \\ 
    \midrule
    Ratio $N_{\mathrm{Ops,LMS}} / N_{\mathrm{Ops,SSC}}$ 
    & \phantom{0}13.82 & \phantom{0}9.67 & 10.27 & \phantom{0}14.88 & \phantom{0}11.36 & \phantom{0}11.72 & \phantom{0}18.91 & \phantom{0}13.71 & \phantom{0}13.68 & \phantom{0}17.52 & \phantom{0}16.88 & \phantom{00}9.9 \\
    \bottomrule                                        
    \end{NiceTabular}
\end{table*}

\subsection{LDPC Code Design}
Since \ac{LDPC} code design is a complex task and beyond the scope of this paper, we consider state-of-the-art standardized \ac{LDPC} codes.
Given that the 5G NR \ac{LDPC} code supports payload lengths only up to $K = 8448$ bits, we instead adopt the CCSDS AR4JA code \cite{bluebook}, which supports $K \in \{1024,\,4096,\,16384\}$ and code rates $R \in \{\nicefrac{1}{2},\,\nicefrac{2}{3},\,\nicefrac{4}{5}\}$.
For even longer block lengths, we use the DVB-S2 \ac{LDPC} code \cite{DVBS2} with $N = 64800$ and the same set of rates.
It is worth noting that, similar to the 5G NR \ac{LDPC} code, the CCSDS code employs punctured \acp{VN} to improve the decoding threshold, albeit at the cost of slightly increased complexity. 
It should be further noted that different code designs exist for \ac{LDPC} codes that optimize performance at lower iteration counts (see \cite{koike2016iterationaware} and references therein). While such designs can improve the performance at lower iteration counts and reduce computational demand, to the best of our knowledge, they have not been standardized yet.

\subsection{Polar Code Design}
Polar codes for \ac{SC} decoding can be optimally designed using \ac{DE} \cite{DensityEvolution}.
For a fair comparison, we construct polar codes with the exact same parameters as the considered \ac{LDPC} codes, with the design \ac{SNR} set for a target \ac{BLER} of $10^{-6}$. 
Since polar codes require block lengths that are powers of two, length matching is performed from the next larger power of two.
We employ shortening as the length-matching technique which is performed from the end of the codeword, either in natural decreasing order or in bit-reverse order \cite{nui2013lengthmatching}, depending on which method yields better performance. Table~\ref{tab:polarlength} lists which variant is used for each parameter combination.

\begin{table}[ht]
\centering
\caption{\footnotesize The used polar code shortening for a BLER of $10^{-6}$.}
\label{tab:polarlength}
\begin{NiceTabular}{c|c c c c}
\toprule
            $R$   & $K = 1024$ & $K = 4096$    & $K = 16384$  & $N = 64800$ \\
\midrule
$\nicefrac{1}{2}$ &    Full &        Full &        Full & Bit-reverse\\
$\nicefrac{2}{3}$ & Natural & Bit-reverse & Bit-reverse &     Natural\\
$\nicefrac{4}{5}$ & Natural &     Natural & Bit-reverse &     Natural\\
\bottomrule
\end{NiceTabular}
\end{table}

\subsection{Single-Iteration Complexity Comparison}
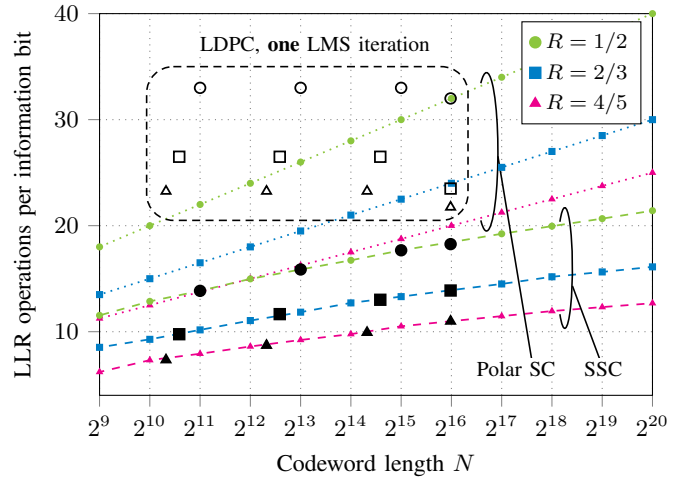
\begin{figure}[ht]
	\centering
    {\begin{tikzpicture}
\begin{axis}[
    width=\linewidth,
    height=0.75\linewidth,
    grid style={dotted,gray},
    ymajorgrids=true,
    xmajorgrids=true,
    xmajorgrids,
    yminorticks=true,
    ymajorgrids,
    tick align=outside,
    tick pos=left,
    label style={font=\small},
    tick label style={font=\small}, 
    xlabel={Codeword length $N$},
    ylabel={LLR operations per information bit},
    xmode=log,
    log basis x=2,
xtick={8,16,32,64,128,256,512,1024,2048,4096,8192,16384,32768,65536,131072,262144,524288,1048576,2097152,4194304,8388608,16777216},
    xticklabels={$2^3$,$2^4$,$2^5$,$2^6$,$2^7$,$2^8$,$2^9$,$2^{10}$,$2^{11}$,$2^{12}$,$2^{13}$,$2^{14}$,$2^{15}$,$2^{16}$,$2^{17}$,$2^{18}$,$2^{19}$,$2^{20}$,$2^{21}$,$2^{22}$,$2^{23}$,$2^{24}$},
    grid=both,
    legend style={only marks, at={(0.98,0.98)}, anchor=north east},
    xmin=512,
    xmax=1048576,
    ymin=4,
    ymax=40,
    every axis plot/.style={
        line width=0.67pt,
        mark size=2pt,
        mark options={solid}
    }
]

\addplot[mark=*, color=R12] coordinates {
    (1, 1)
};
\addplot[mark=square*, color=R23] coordinates {
    (1, 1)
};
\addplot[mark=triangle*, color=R45] coordinates {
    (1, 1)
};

\addplot[mark=*, dotted, color=R12, mark size=1.pt] coordinates {
(8, 6.0)
(16, 8.0)
(32, 10.0)
(64, 12.0)
(128, 14.0)
(256, 16.0)
(512, 18.0)
(1024, 20.0)
(2048, 22.0)
(4096, 24.0)
(8192, 26.0)
(16384, 28.0)
(32768, 30.0)
(65536, 32.0)
(131072, 34.0)
(262144, 36.0)
(524288, 38.0)
(1048576, 40.0)
(2097152, 42.0)
(4194304, 44.0)
(8388608, 46.0)
(16777216, 48.0)
};

\addplot[mark=square*, dotted, color=R23, mark size=1.pt] coordinates {
(8, 4.5)
(16, 6.0)
(32, 7.5)
(64, 9.0)
(128, 10.5)
(256, 12.0)
(512, 13.5)
(1024, 15.0)
(2048, 16.5)
(4096, 18.0)
(8192, 19.5)
(16384, 21.0)
(32768, 22.5)
(65536, 24.0)
(131072, 25.5)
(262144, 27.0)
(524288, 28.5)
(1048576, 30.0)
(2097152, 31.5)
(4194304, 33.0)
(8388608, 34.5)
(16777216, 36.0)
};

\addplot[mark=triangle*, dotted, color=R45, mark size=1.pt] coordinates {
(8, 3.75)
(16, 5.0)
(32, 6.25)
(64, 7.5)
(128, 8.75)
(256, 10.0)
(512, 11.25)
(1024, 12.5)
(2048, 13.75)
(4096, 15.0)
(8192, 16.25)
(16384, 17.5)
(32768, 18.75)
(65536, 20.0)
(131072, 21.25)
(262144, 22.5)
(524288, 23.75)
(1048576, 25.0)
(2097152, 26.25)
(4194304, 27.5)
(8388608, 28.75)
(16777216, 30.0)
};

\addplot[mark=*, dashed, color=R12, mark size=1.pt] coordinates {
(8, 3.5)
(16, 3.75)
(32, 6.625)
(64, 7.0625)
(128, 9.25)
(256, 10.109375)
(512, 11.5390625)
(1024, 12.86328125)
(2048, 13.85351562)
(4096, 14.98242188)
(8192, 15.86865234)
(16384, 16.73510742)
(32768, 17.67932129)
(65536, 18.45031738)
(131072, 19.23464966)
(262144, 19.95457458)
(524288, 20.66485596)
(1048576, 21.41268921)
(2097152, 21.99247169)
(4194304, 22.66471863)
(8388608, 23.1533947)
(16777216, 23.61662459)
};

\addplot[mark=square*, dashed, color=R23, mark size=1.pt] coordinates {
(8, 2.0)
(16, 3.36363636)
(32, 3.72727273)
(64, 5.55813953)
(128, 6.18604651)
(256, 7.49122807)
(512, 8.52631579)
(1024, 9.28111274)
(2048, 10.1727672)
(4096, 11.05199561)
(8192, 11.84328085)
(16384, 12.71765998)
(32768, 13.3092557)
(65536, 13.921929)
(131072, 14.50415417)
(262144, 15.18087353)
(524288, 15.63757775)
(1048576, 16.11683411)
(2097152, 16.59324999)
(4194304, 17.0666604)
(8388608, 17.44499952)
(16777216, 17.85251615)
};

\addplot[mark=triangle*, dashed, color=R45, mark size=1.pt] coordinates {
(8, 1.0)
(16, 2.07692308)
(32, 3.23076923)
(64, 3.38461538)
(128, 4.78640777)
(256, 5.63414634)
(512, 6.2)
(1024, 7.31707317)
(2048, 7.92007322)
(4096, 8.60573695)
(8192, 9.22093378)
(16384, 9.77220018)
(32768, 10.51089071)
(65536, 10.99671937)
(131072, 11.46863377)
(262144, 11.94537374)
(524288, 12.32138063)
(1048576, 12.68654878)
(2097152, 13.10511515)
(4194304, 13.48044014)
(8388608, 13.8009433)
(16777216, 14.15593089)
};

\addplot[mark=*, color=black] coordinates {
    (8192, 15.86865234375)
};
\addplot[mark=square*, color=black] coordinates {
    (6144, 11.657958984375)
};
\addplot[mark=triangle*, color=black] coordinates {
    (5120, 8.7099609375)
};

\addplot[mark=*, color=black] coordinates {
    (2048, 13.853515625)
};
\addplot[mark=square*, color=black] coordinates {
    (1536, 9.763671875)
};
\addplot[mark=triangle*, color=black] coordinates {
    (1280, 7.34375)
};

\addplot[mark=*, color=black] coordinates {
    (32768, 17.6793212890625)
};
\addplot[mark=square*, color=black] coordinates {
    (24576, 13.01312255859375)
};
\addplot[mark=triangle*, color=black] coordinates {
    (20480, 9.9290771484375)
};

\addplot[mark=*, color=black] coordinates {
    (64800, 18.259814814814813)
};
\addplot[mark=square*, color=black] coordinates {
    (64800, 13.884722222222223)
};
\addplot[mark=triangle*, color=black] coordinates {
    (64800, 10.989043209876543)
};

\addplot[mark=o, color=black] coordinates {
    (2048, 33.0)
};

\addplot[mark=square, color=black] coordinates {
    (1536, 26.5)
};

\addplot[mark=triangle, color=black] coordinates {
    (1280, 23.25)
};

\addplot[mark=o, color=black] coordinates {
    (8192, 33.0)
};

\addplot[mark=square, color=black] coordinates {
    (6144, 26.5)
};

\addplot[mark=triangle, color=black] coordinates {
    (5120, 23.25)
};

\addplot[mark=o, color=black] coordinates {
    (32768, 33.0)
};

\addplot[mark=square, color=black] coordinates {
    (24576, 26.5)
};

\addplot[mark=triangle, color=black] coordinates {
    (20480, 23.25)
};

\addplot[mark=o, color=black] coordinates {
    (64800, 32)
};

\addplot[mark=square, color=black] coordinates {
    (64800, 23.5)
};

\addplot[mark=triangle, color=black] coordinates {
    (64800, 21.75)
};

\draw[line width=0.6pt] (axis cs: 292544, 11.65) arc(-130:130:0.1cm and 0.8cm);
\coordinate (arc1) at ($(axis cs: 292544, 11.65)+(0.165cm,0.375cm)$);
\draw[line width=0.6pt] (arc1) -- (axis cs: 532544, 6.5);
\node[rectangle, align=center, inner sep=1pt, fill=white] at (axis cs: 532544, 6.5){\footnotesize SSC};

\draw[line width=0.6pt] (axis cs: 98304, 20.5) arc(-120:120:0.15cm and 1.05cm);
\coordinate (arc2) at ($(axis cs: 98304, 20.5)+(0.225cm,0.85cm)$);
\draw[line width=0.6pt] (arc2) -- (axis cs: 202544, 6.5);
\node[rectangle, align=center, inner sep=1pt, fill=white] at (axis cs: 160000, 6.5){\footnotesize Polar SC};

\draw[line width=0.6pt, rounded corners=10pt, densely dashed] (axis cs: 980, 35) -- (axis cs: 82384, 35) -- (axis cs: 82384, 20.5) -- (axis cs: 980, 20.5) -- cycle;
\node[rectangle, align=center, inner sep=1pt, fill=white] at (axis cs: 10000, 37){\footnotesize LDPC, \textbf{one} \ac{LMS} iteration};

\legend{\footnotesize $R = 1/2$, \footnotesize $R = 2/3$, \footnotesize $R = 4/5$}

\end{axis}
\end{tikzpicture}}
	\caption{\footnotesize Decoding complexity comparison: A single \ac{LDPC} \ac{LMS} iteration (hollow markers) vs. full polar \ac{SC}/\ac{SSC} decoding (solid markers).}
    \label{fig:complexityperinfobit}
    \vspace{-1.8em}
\end{figure}

To get an intuition for the scaling of the complexity, we compute the number of \ac{LLR} operations for polar \ac{SC}/\ac{SSC} and a single iteration of \ac{LDPC} \ac{LMS} decoding for each of the proposed codes.
We plot these values normalized to a single information bit, i.e., $N_\mathrm{Ops}/K$ in Fig.~\ref{fig:complexityperinfobit}.
The curves for polar codes are computed for \ac{DE}-based code design. The black markers show the complexity for each of the explicit parameters, including shortening.
We observe that while for $N=64800$, \ac{SC} and \ac{LMS} show similar complexity, full \ac{SSC} decoding of polar codes is still significantly less complex (approx. half) than a single \ac{LMS} iteration of \ac{LDPC} codes.

\subsection{Performance-Matched Complexity Comparison}
\begin{figure*}[htp]
	\centering
	\resizebox{\linewidth}{!}{\input{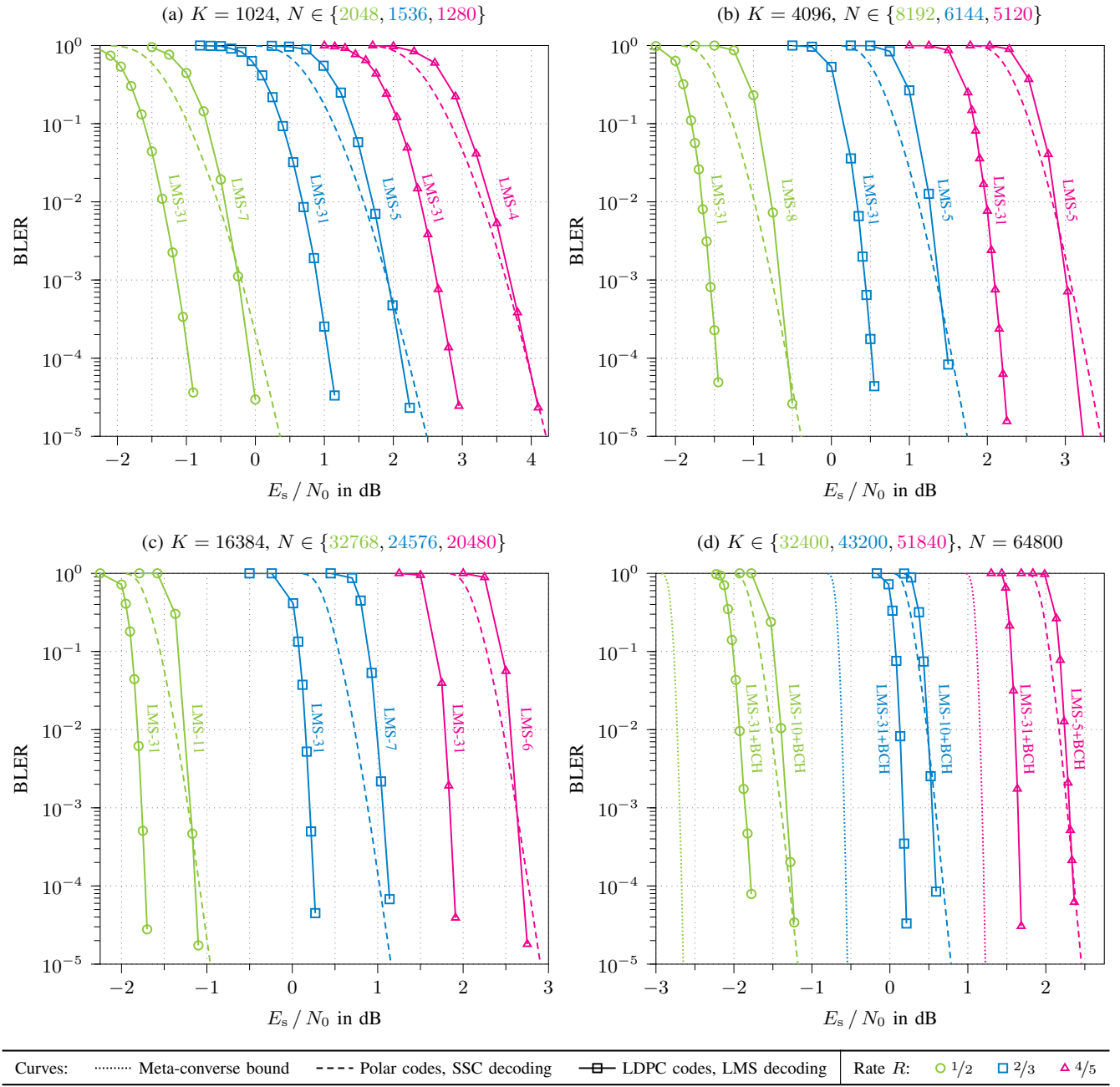}}
	\caption{\footnotesize \ac{BLER} comparison of polar codes under \ac{SC} decoding (dashed curves) and \ac{LMS} decoding of \ac{LDPC} codes (solid curves). The polar codes demonstrate competitive performance compared to the \ac{LDPC} codes with low iteration counts. \ac{LDPC} codes with a sufficient number of iterations outperform polar codes.}
	\label{fig:bler}
	\vspace{-1.45em}
\end{figure*}

Since a single iteration is insufficient to achieve any meaningful error-correction performance (\ac{BLER} $\approx 1$), we now focus on comparing the computational complexity for a larger number of iterations.
To this end, we tune the number of iterations of the \ac{LMS} decoder to match the \ac{BLER} performance of polar codes under \ac{SC} decoding.
Fig.~\ref{fig:bler} presents the error-correction performance across all considered code parameters, with the number of \ac{LMS} iterations indicated for each curve.
Moreover, we show the saddle point approximations of the \ac{PPV} meta-converse bound \cite{SaddlePointApproxMC}, indicating a finite-length performance limit for any code.
As we can see, the \ac{LDPC} codes can match the performance of polar codes at \ac{BLER} of $10^{-3}$ using 5 to 11 iterations.
We observe that longer and lower-rate \ac{LDPC} codes generally require more iterations than their shorter or higher-rate counterparts.
While the \ac{BLER} curves of polar codes begin to bend at lower \ac{SNR} values, those of \ac{LDPC} codes are steeper, indicating a sharper transition in the waterfall region.
As expected, allowing for even larger number of iterations enables \ac{LDPC} codes to significantly outperform polar codes.

For a fair complexity comparison, we consider the average number of iterations until convergence, using early stopping.
However, due to the steep nature of the \ac{LDPC} \ac{BLER} curves, early stopping has limited impact—most correctly decoded frames in the waterfall region still require the full $N_{\mathrm{it}}$ iterations.
Consequently, the average number of iterations is nearly equal to the maximum.
Table~\ref{tab:complexity} summarizes the number of \ac{LLR} operations per information bit, i.e., $N_\mathrm{Ops}/K$, for \ac{SSC} decoding of polar codes and \ac{LMS} decoding of \ac{LDPC} codes, and their ratio $N_{\mathrm{Ops,LMS}} / N_{\mathrm{Ops,SSC}}$.
Table~\ref{tab:complexity} reveals that decoding \ac{LDPC} codes requires at least approximately ten times more computational complexity than polar codes. This higher complexity stems from both the costlier per-iteration operations and the need for multiple iterations to match polar code performance. The difference is especially pronounced at lower code rates like $R = \nicefrac{1}{2}$, where polar codes naturally perform better. For \ac{LDPC} codes, the high iteration count at these lower rates and the limited effectiveness of early stopping further increases computational demands. In extreme cases, \ac{LDPC} decoders require up to $19$ times more operations than their polar code counterparts.

\section{Summary and Discussion}
In this work, we compared soft-input decoding of long block codes, focusing on polar codes under \ac{SC} decoding and \ac{LDPC} codes under iterative \ac{BP}-based decoding.
Our results indicate that a single \ac{BP} iteration requires more operations than full \ac{SSC} decoding, making it hard for \ac{LDPC} codes to match the performance of polar codes \textit{at similar computational complexity}---even with early stopping.
To achieve comparable error-correction performance, \ac{LDPC} codes need higher computational effort.
Polar codes demonstrate particular strength at long block lengths and low to medium code rates, offering attractive performance in these regimes.
However, as the number of decoding iterations is increased, \ac{LDPC} codes do still outperform polar codes and exhibit a steeper slope in the \ac{BLER} chart.
While \ac{BP}-based decoding of \ac{LDPC} codes allows flexible adjustment of complexity via the number of iterations, enhancements to polar decoding---such as list, flip, or perturbation decoding---typically incur disproportionately high implementation costs \cite{Kestel2023URLLC,ercan2018flip,yang2025perturbation}.
At practical error rates for single-shot, end-to-end transmission, the steep waterfall behavior of \ac{LDPC} codes eventually surpasses the shallower \ac{BLER} curve of polar codes under \ac{SC} decoding. 
Nonetheless, in mobile data-channel applications, where \ac{HARQ} adapts coding rates to varying channel conditions and operates in the $1\%$–$10\%$ \ac{BLER} range, the early bending of the polar curve can be advantageous.

While our complexity evaluation based on the number of \ac{LLR} operations may only provide a rough estimate of real-world hardware complexity and energy efficiency, the observed differences are substantial enough to suggest a similar trend in \ac{ASIC} implementations.
We encourage both academia and industry to validate these insights through \ac{ASIC} synthesis and to report post place-and-route implementation results for decoders of long codes.

\bibliographystyle{IEEEtran}
\bibliography{references.bib}
\end{NoHyper}
\end{document}